\documentclass[12pt]{article}

\usepackage[english]{babel}
\usepackage{graphicx,rotating,epsfig}
\usepackage{a4p}
\usepackage{cite,mcite}

\def\GeV  {\ensuremath{\mathrm{ Ge\kern -0.1em V } }}
\def\GeVc2{\ensuremath{\mathrm{ Ge\kern -0.1em V }\kern -0.2em /c^2 }}
\newcommand{\MT}{\ensuremath{M_{\mathrm{top}}}}
\newcommand{\MW}{\ensuremath{M_{\mathrm{ W }}}}
\newcommand{\Pt}{\ensuremath{t}}
\newcommand{\Ptt}{\ensuremath{\Pt\bar\Pt}}

\begin{document}

\begin{center}
{\LARGE FERMI NATIONAL ACCELERATOR LABORATORY}
\end{center}

\begin{flushright}
       TEVEWWG/top 2004/01\\
       CDF Note 6955  \\
       D\O\ Note 4417 \\
       hep-ex/0404010 \\
       {\bf 8th April 2004}
\end{flushright}

\vskip 1cm

\begin{center}
{\Huge \bf Combination of CDF and D\O\ Results \\[3mm]
                  on the Top-Quark Mass}
\vskip 1cm
{\Large
The CDF Collaboration, the D\O\ Collaboration, and \\[1mm]
the Tevatron Electroweak Working Group\footnote{
WWW access at {\tt http://tevewwg.fnal.gov}\\
The members of the TEVEWWG who contributed significantly to the
analysis described in this note are:
P.~Azzi (azzi@fnal.gov),                %
E.~Barberis (barberis@fnal.gov),        %
L.~Demortier (luc@fnal.gov),            %
M.~W.~Gr\"unewald (mwg@fnal.gov),       %
A.~Juste (juste@fnal.gov),              %
B.~Klima (klima@fnal.gov),              %
J.~Konigsberg (konigsberg@fnal.gov),    %
A.~Quadt (quadt@fnal.gov),              %
M.~Strovink (strovink@lbl.gov),         %
E.~Thomson (thomsone@fnal.gov).         %
}
}

\vskip 1cm

{\bf Abstract}

\end{center}

{%
  
  The results on the measurements of the top-quark mass, based on the
  data collected by the Tevatron experiments CDF and D\O\ at Fermilab
  during Run I from 1992 to 1996, are summarized. The combination of
  the published results, taking correlated uncertainties properly into
  account, is presented.  The resulting world average for the mass of
  the top quark is: $\MT=178.0\pm4.3~\GeVc2$, where the total error of
  consists of a statistical part of $2.7~\GeVc2$ and a systematic part
  of $3.3~\GeVc2$.

}

\vfill

\section{Introduction}

The experiments CDF and D\O, taking data at the Tevatron
proton-antiproton collider located at the Fermi National Accelerator
Laboratory, have published several direct experimental measurements of
the pole mass, $\MT$, of the top quark
$\Pt$~\cite{Mtop1-CDF-di-l-PRLa, Mtop1-CDF-di-l-PRLb,
Mtop1-CDF-di-l-PRLb-E, Mtop1-D0-di-l-PRL, Mtop1-D0-di-l-PRD,
Mtop1-CDF-l+j-PRL, Mtop1-CDF-l+j-PRD, Mtop1-D0-l+j-old-PRL,
Mtop1-D0-l+j-old-PRD, Mtop1-D0-l+j-new, Mtop1-CDF-all-j-PRL}. The
measurements are based on Run I data (1992-1996) and utilize all decay
topologies arising in $\Ptt$ production given by the leptonic or
hadronic decay of the W boson occurring in top-quark decay: the
di-lepton channel (di-l)~\cite{Mtop1-CDF-di-l-PRLa,
Mtop1-CDF-di-l-PRLb, Mtop1-CDF-di-l-PRLb-E, Mtop1-D0-di-l-PRL,
Mtop1-D0-di-l-PRD}, the lepton+jets channel
(l+j)\cite{Mtop1-CDF-l+j-PRL, Mtop1-CDF-l+j-PRD, Mtop1-D0-l+j-old-PRL,
Mtop1-D0-l+j-old-PRD, Mtop1-D0-l+j-new}, and the all-jets channel
(all-j) analysed by CDF only ~\cite{Mtop1-CDF-all-j-PRL}.  The
lepton+jets channel yields the most precise determination of
$\MT$. The recently published new measurement in this channel by the
D\O\ collaboration~\cite{Mtop1-D0-l+j-new} is based on a powerful
analysis technique yielding a much smaller measurement uncertainty.

This note reports on the combination of the most recent final and
comprehensive mass measurements in each channel by CDF and
D\O~\cite{Mtop1-CDF-di-l-PRLb, Mtop1-CDF-di-l-PRLb-E,
Mtop1-D0-di-l-PRD, Mtop1-CDF-l+j-PRD, Mtop1-D0-l+j-new,
Mtop1-CDF-all-j-PRL}.  The combination takes into account the
statistical and systematic uncertainties as well as the correlations
between systematic uncertainties, and replaces the previous
combination~\cite{TM-2084}.  The new D\O\
measurement~\cite{Mtop1-D0-l+j-new} is the single most precise
top-quark mass measurement and has the largest weight in this new
combination.

\section{Measurements}

The five measurements of $\MT$ to be combined are listed in
Table~\ref{tab:inputs}.  Besides central values and statistical
uncertainties, the systematic errors arising from various sources are
reported. In order of decreasing importance, the systematic error
sources are:
\begin{itemize}
\item Jet energy scale (JES): The systematics for jet energy scale
include the uncertainties on the absolute jet energy corrections,
calorimeter stability, underlying event and relative jet energy
corrections.

\item Model for signal (signal): The systematics for the signal model
include initial and final state radiation effects, b-tagging bias,
dependence upon parton distribution functions as well as variations in
$\Lambda_{\mathrm{QCD}}$.

\item Model for background (BG): The background model includes
estimates of the effect of setting $Q^2=\langle p_t\rangle^2$ instead
of $Q^2=\MW^2$ in VECBOS~\cite{VECBOS} simulations of W+jets
production, the use of ISAJET~\cite{ISAJET} fragmentation
instead of HERWIG~\cite{HERWIG5} fragmentation as well as the effect
of varying the background fraction attributed to QCD.

\item Uranium noise and multiple interactions (UN/MI): This
uncertainty includes uncertainties arising from uranium noise in the
D\O\ calorimeter and from multiple interactions overlapping signal
events.  CDF includes the systematic uncertainty due to multiple
interactions in the JES contribution.

\item Method for mass fitting (fit): This systematic uncertainty takes
into account the finite sizes of Monte Carlo samples used for fitting,
impact of jet permutations, and other fitting biases.  In the CDF
lepton+jets analysis, the systematic uncertainty due to finite Monte
Carlo statistics is included in the statistical uncertainty.

\item Monte Carlo generator (MC): The systematic uncertainty on the
Monte Carlo generator provides an estimate of the sensitivity to the
simulated physics model by comparing HERWIG to PYTHIA~\cite{PYTHIA4,
*PYTHIA5} or to ISAJET.  In the D\O\ analyses, the systematic
uncertainty associated with the comparison of HERWIG to ISAJET is
included in the signal model uncertainty.

\end{itemize}
For each measurement, the individual error contributions are combined
in quadrature.

\begin{table}[h]
\begin{center}
\renewcommand{\arraystretch}{1.30}
\begin{tabular}{|l||rrr|rr|}
\hline
Run I  &CDF l+j & CDF di-l & CDF all-j & D\O\ l+j & D\O\ di-l \\
\hline
\hline
Result &  176.1 &    167.4 &     186.0 &    180.1 &     168.4 \\
\hline
\hline
Stat. &     5.1 &     10.3 &      10.0 &      3.6 &      12.3 \\
\hline
\hline
JES   &     4.4 &      3.8 &       5.0 &      3.3 &       2.4 \\
Signal&     2.6 &      2.8 &       1.8 &      1.1 &       1.8 \\
BG    &     1.3 &      0.3 &       1.7 &      1.0 &       1.1 \\
UN/MI &     0.0 &      0.0 &       0.0 &      1.3 &       1.3 \\
Fit   &     0.0 &      0.7 &       0.6 &      0.6 &       1.1 \\
MC    &     0.1 &      0.6 &       0.8 &      0.0 &       0.0 \\
\hline
Syst. &     5.3 &      4.8 &       5.7 &      3.9 &       3.6 \\
\hline
\hline
Total &     7.3 &     11.4 &      11.5 &      5.3 &      12.8 \\
\hline
\end{tabular}
\end{center}
\caption[Input measurements]{Summary of the five measurements of $\MT$
performed by CDF and D\O. All numbers are in $\GeVc2$. For each
measurement, the corresponding column lists experiment and channel,
central value and contributions to the total error, namely statistical
error and systematic errors arising from various sources defined in
the text. Overall systematic errors and total errors are obtained by
combining individual errors in quadrature.}
\label{tab:inputs}
\end{table}

\section{Combination}

In the combination, the error contributions arising from different
sources are uncorrelated between measurements.  The correlations of
error contributions arising from the same source are as follows:
\begin{itemize} 
\item uncorrelated: statistical error, fit error;
\item 100\% correlated within each experiment: JES error, UN/MI error;
\item 100\% correlated within each channel: BG error;
\item 100\% correlated between all measurements: signal error, MC error.
\end{itemize}
The resulting matrix of global correlation coefficients is listed in
Table~\ref{tab:correl}.

\begin{table}[h]
\begin{center}
\renewcommand{\arraystretch}{1.30}
\begin{tabular}{|l||rrr|rr|}
\hline
Run I     & CDF l+j & CDF di-l & CDF all-j & D\O\ l+j & D\O\ di-l \\
\hline
\hline
CDF l+j   &    1.00 &          &           &          &           \\
CDF di-l  &    0.29 &     1.00 &           &          &           \\
CDF all-j &    0.32 &     0.19 &      1.00 &          &           \\
\hline
D\O\ l+j  &    0.11 &     0.05 &      0.03 &     1.00 &           \\
D\O\ di-l &    0.05 &     0.04 &      0.02 &     0.17 &      1.00 \\
\hline
\end{tabular}
\end{center}
\caption[Global correlations between input measurements]{Matrix of
global correlation coefficients between the five measurements of
$\MT$.}
\label{tab:correl}
\end{table}

The five measurements are combined using two independent programs: one
which has already been used for the previous
combination~\cite{TM-2084}, and one implementing a numerical $\chi^2$
minimisation as well as the analytic BLUE method~\cite{Lyons:1988,
Valassi:2003}. The three methods used are mathematically equivalent,
and give identical results for the combination. In addition, the BLUE
method yields the decomposition of the error on the average in terms
of the error categories specified for the input
measurements~\cite{Valassi:2003}.

\section{Results}

The combined value for the top-quark mass is:
\begin{eqnarray}
\MT & = & 178.0\pm4.3~\GeVc2\,,
\end{eqnarray}
where the total error of $4.3~\GeVc2$ contains the following
components: a statistical error of $2.7~\GeVc2$; and systematic error
contributions of: JES $2.6~\GeVc2$, signal $1.6~\GeVc2$, background
$0.88~\GeVc2$, UN/MI $0.83~\GeVc2$, fit $0.35~\GeVc2$, and MC
$0.12~\GeVc2$, for a total systematic error of $3.3~\GeVc2$.

\begin{figure}[t]
\begin{center}
\includegraphics[width=0.8\textwidth]{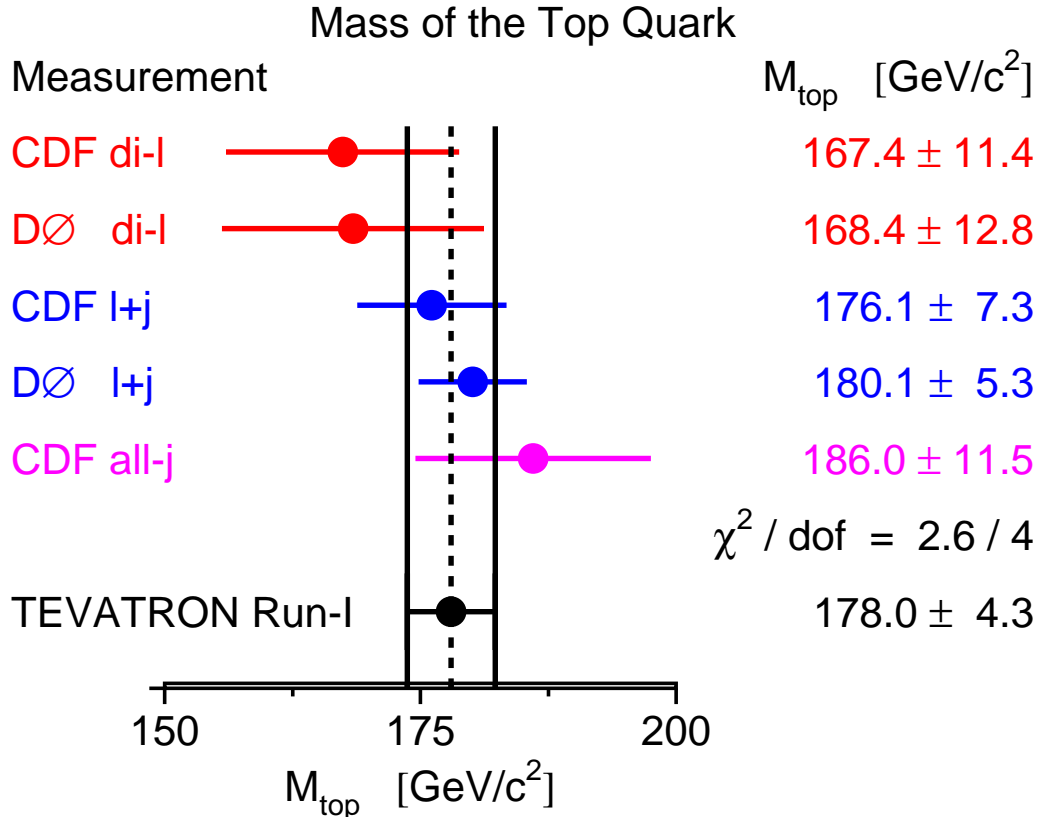}
\end{center}
\caption[Comparison of the measurements of the top-quark mass]
{Comparison of the measurements of the top-quark mass and their
average.}
\label{fig:mtop-bar-chart} 
\end{figure}

The $\chi^2$ of this average is 2.6 for 4 degrees of freedom,
corresponding to a probability of 63\%, showing that all measurements
are in good agreement with each other which can also be seen in
Figure~\ref{fig:mtop-bar-chart}.  The pull of each measurement with
respect to the average and the weight of each measurement in the
average are reported in Table~\ref{tab:stat}.

\begin{table}[t]
\begin{center}
\vskip 7mm
\renewcommand{\arraystretch}{1.30}
\begin{tabular}{|l||rrr|rr|}
\hline
Run I     & CDF l+j & CDF di-l & CDF all-j & D\O\ l+j & D\O\ di-l \\
\hline
\hline
Pull      & $-0.32$ &  $-1.01$ &   $+0.75$ &  $+0.66$ &   $-0.80$ \\
\hline
Weight    &  0.222  &   0.071  &    0.069  &   0.578  &    0.060  \\
\hline
\end{tabular}
\end{center}
\caption[Pull and weight of each measurement]{Pull and weight of each
  measurement in the average.}
\label{tab:stat} 
\end{table}

\section{Summary}

An updated combination of the five published measurements of $\MT$
from CDF and D\O\ based on Run I data is presented. Taking into
account statistical and systematic errors including their
correlations, the Tevatron and thus the world-average result is:
$\MT=178.0\pm4.3~\GeVc2$.  The mass of the top quark is now known with
an accuracy of 2.4\%.

\bibliographystyle{tevewwg}
\bibliography{run1mtop}

\end{document}